\documentclass[11pt,a4paper]{article}
\usepackage{amsfonts}
\usepackage{graphicx}
\usepackage{amsmath}
\usepackage{hyperref}
\usepackage{enumerate}

\RequirePackage{mathrsfs} \RequirePackage[sc]{mathpazo}
\RequirePackage{wasysym} \RequirePackage{setspace}
\makeatletter \@addtoreset{equation}{section}

\newcommand{\be}{\begin{equation}}
\newcommand{\ee}{\end{equation}}
\newcommand{\bea}{\begin{eqnarray}}
\newcommand{\eea}{\end{eqnarray}}
\begin{document}

\title{\textbf{\ On Fractional Quantum Hall Solitons and  }\\
\textbf{ Chern-Simons Quiver Gauge  Theories} }
\author{Adil  Belhaj\thanks{%
belhaj@unizar.es}\hspace*{-15pt} \\
\\
{\small Centre of Physics and Mathematics, CPM-CNESTEN, Rabat, Morocco} \\
{\small Groupement National de Physique des Hautes Energies, Rabat,
Morocco }} \maketitle

\begin{abstract}
We investigate  a class of hierarchical  multiple layers of
fractional quantum Hall soliton (FQHS)  systems from Chern-Simons
quivers embedded in   M-theory on   the cotangent on a 2-dimensional
complex toric variety $\bf V^2$, which  is dual to type IIA
superstring on  a 3-dimensional complex manifold $\bf  {CP}^1\times
V^2$ fibered over a real line $\mathbb{R}$.  Based on M-theory/Type
IIA duality, FQHS systems can be derived from wrapped D4-branes on
2-cycles in $\bf {CP}^1\times V^2$ type IIA geometry.  In this
realization, the magnetic source can be identified  with  gauge
fields  obtained from the decomposition of the R-R 3-form on a
generic combination  of  2-cycles. Using   type IIA D-brane flux
data, we compute the filling factors  for   models relying on $\bf
{CP}^2$ and the zeroth  Hirzebruch surface.
\end{abstract}

\newpage

\section{Introduction}

Three dimensional Quantum Hall Systems (QHS) of condensed matter
physics can be  modeled using D-branes of type II superstrings
\cite{BBST,FLRT,BJLL}. The first  ten dimensional superstring
picture of Quantum Hall Effect (QHE) in (1+2)-dimensions has been
given in terms of a stack of $K$ D6-branes, a spherical D2-brane
as well as dissolved D0-branes in D2 and a stack of $N$ F-strings
stretching between D2 and D6-branes \cite{BBST}.  The F-strings ending
on the D2-brane have an interpretation in terms of the fractional
quantum Hall particles  and they are charged under
the U$(1)$ world-volume gauge field associated with the D0-branes. These solitonic objects
 behave as magnetic flux quanta disolved in the D2-brane
world-volume on which the QHS system  reside.  This string picture
of QHE in (1+2)-dimensions has been extended to the compactification
of type IIA superstring theory on  the  K3 surface  with
singularities classified by Dynkin diagrams\cite{BS,BFSS,BEFKSS}.

Alternatively,  some  efforts have been devoted to study fractional
quantum Hall solitons (FQHS)  in
 connections  with (2+1)-dimensional Chern-Simons (CS) theory
constructed by Aharony, Bergman, Jafferis and Maldacena. Recall that, ABJM
theory  is a 3-dimensional
  $N=6$ CS  quiver  with  $\mbox{U(N)}_k \times  \mbox{U(N)}_{-k}$  gauge symmetry  proposed to be dual to M-theory
  propagating on $AdS_4\times S^7/Z_k$, with an appropriate amount of fluxes, or type IIA superstring on
    $AdS_4\times \bf CP^3$ for large  $k$ and $N$ with $k\ge N$ in the weakly interacting regime  \cite{ABJM}.
    In the decoupling limit, the corresponding  three dimensional conformal field theory (CFT$_3$)  is obtained from the physics of  the  multiple
     M2-branes placed at the orbifold  space $ C^4/Z_k$.
It has been shown that QHE  can be realized  on the world-volume
action of the M5-brane filling AdS$_3$ inside AdS$_4$\cite{F}. The
model has been derived from $d = 3$ flavored ABJM theory with the CS
levels $(1,-1)$.     FQHE systems can be also embedded in ABJM theory
by   adding fractional type IIA D-branes \cite{HLT}.

Recently, a possible extension of FQHS in ABJM theory  has been
given using  type  IIA dual geometry  considered  as  the  blown up
of $\bf
 {CP}^3$ by four-cycles which are isomorphic  to  $\bf  {CP}^2$. Based on   D6-branes wrapped 4-cycles and
interacting with  the R-R gauge fields in    ABJM-like geometry,
 hierarchical  multiple layers  of FQHS systems  have been given  in  \cite{B}.

Exended constructions  can  be  characterized by a vector $q_i$ and a real, symmetric and invertible symmetric matrix $K_{ij}$. These parameters classify various fractional quantum Hall (FQH) states.  The choice of $K_{ij}$   play important role
in the  embedding of  CS  in  type II superstrings and M-theory compactified on deformed singular Calabi-Yau manifolds. In particular,
the matrix $K_{ij}$  can be  identified with the  intersection numbers of compact cycles in the internal space.
These numbers  are, up to some details, the opposite of the  Cartan matrices  of Lie algebras.
This link  may lead to a nice correspondence between  FQHS models   and  2-cycles involved
in the deformation of  toric  singularities. More specifically, to each simple 2-cycle, we associate
a FQHS model.

In this work we discuss such  CS  realizations of  QHS in 1+2 dimensions  from M-theory  on a real eight
dimensional manifold. The manifold  is realized explicitly as the
cotangent bundle over a 2-dimensional complex toric variety $\bf
V^2$.  Up some details, the obtained  models   can be  compraed to Gaiotto-Witten theory   describing $ N=4$  CS  superconformal theories. This connection can be
obtained by imposing constraints on the possible  values on CS levels satisfying the fundamental identity \cite{GW}.\\ Since   $q_i$ and  $K_{ij}$
are connected via the filling factor, the embedding of CS in the string theory gives a toric  geometric calculation  for  the filling factor.
More precisely,  starting
 from given  two toric varieties  describing  quantum Hall systems and using
                            CS gauge modeling a la Susskind, we determine the filling factors.
In particular,  based on M-theory/Type IIA duality, we give first
  Chern-Simons type theories describing 3-dimensional FQHS systems
using D4-branes wrapping  2-cycles  and interacting with the
 R-R gauge fields living on $\bf  {CP}^1\times
V^2$   type IIA  dual geometry.  This class of QHS can be considerd as a possible extension  of the works given  discusing the QHE in ABJM and its generalization \cite{F,HLT,B}.  This allows a geometric interpretation of
filling factors in terms of the  intersections between 2-cycles in  $H^2({\bf
V^2},Z)$.
 The matrix $K_{ij}$ can be identified with  the intersection
matrix for curves embedded in $\bf  {CP}^1\times
V^2$.  Then, we present  explicit examples  in order to illustrate the general idea. In particular, we discuss the case of  $\bf  {CP}^2$   and the  Hirzebruch surfaces.

The organization of this  work  is as follows. In the next section,
we  study FQHS systems in ABJM-like  theory  using D6-branes
wrraping 4-cycles in $\bf {CP}^3$. In section 3, we derive  a class
of FQHS in (1+2)-dimensions using M-theory/Type IIA duality. In
particular, we discuss  FQHS  model based on $\bf  {CP}^2$, then we
 extend the analysis to the Hirzebruch surfaces in section 4. The
last section is devoted to our conclusion.

\section{ QHS  in  ABJM  theory}
We start this section  by recalling that  the quantum Hall states are characterized by
the filling factor $\nu$ describing  the ratio  between the
electronic density and the flux density.  When $\nu$ is a fractional
value,   it is called fractional quantum Hall effect (FQHE) for interacting
electron systems. The  first proposed series of the fractional quantum states was given by
Laughlin and they are characterized by the filling factor $\nu _{L}=\frac{1}{%
k}$ where $k$ is an even integer for a boson electron and an odd
integer for a fermionic electron \cite{L}. At low energy, this model
can be described by a 3-dimensional U(1) Chern-Simons theory coupled
to an external electromagnetic field $\tilde{A}$ with the following
effective action
\begin{equation}
S_{CS}=-\frac{k}{4\pi}\int_{\mathbb{R}^{1,2}}A\wedge
dA+\int_{\mathbb{R}^{1,2}}\frac{q}{2\pi}{\tilde{A}}\wedge dA\label{sc},
\end{equation}
where $A$ is the dynamical gauge field, $\tilde{A}$  is  an external
electromagnetic field, and where $q$ is the charge of the electron
\cite{W,WZ}.
 Extended models  are characterized by  a vector $q_i$  and   a real, symmetric and invertible symmetric  matrix
$K_{ij}$. These parameters   play an important role in quiver
construction of FQHE embedded  in  type II superstrings
\cite{BS,BEFKSS} and M-theory compactified on 8-dimensional
manifolds \cite{BFSS}.
 Following the Susskind approach for abelian field theory, these models  can be  described   by the following action
\begin{equation}%
\begin{tabular}
[c]{ll}%
$S\sim\frac{1}{4\pi}\int_{\mathbb{R}^{1,2}} K_{ij}A^{i}\wedge
dA^{j}+2\int_{\mathbb{R}^{1,2}} q_{i}\tilde{A}\wedge dA^{i}.$ &
\end{tabular}
\label{hd}%
\end{equation}  The external
gauge field  $\tilde{A}$
couples  now to each   current $\star dA^i$ with charge strengths  $eq_i$. The $K_{ij%
}$\ matrix and the $q_{i}$  charge vector in this effective field
action suggests  some physical concepts. Following the
Wen-Zee model \cite{WZ}, $K_{ij}$ and $q_{i}$ are interpreted as
order parameters and  classify the various QHS states. Integrating
over the all abelian  gauge fields $A^{i}$, one gets  the formulae
for the filling factor of the system
\begin{equation}%
\begin{tabular}
[c]{ll}%
$\nu=q_{i}K_{ij}^{-1}q_{j}.$ &
\end{tabular}
\label{factor}%
\end{equation}
In what follows, we see that    such  CS quivers
 describing FQHS  can be embedded  either  in  ABJM theory  with
$\mbox{U}(N)_k\times \mbox{U}(N)_{-k}$ gauge symmetry or  more
generally in  toric CS quiver gauge theories. For simplicity,
we first
     consider the case   of   the U(1)$_k$ CS gauge theory in ABJM theory.  Indeed, this model   can be obtained from
a D6-brane   wrapping   a  four-cycle class $[C_4]$
in $H^4( {\bf CP^3},Z)$ which is one-dimensional.  On the gauge
theory side of ABJM, the gauge symmetry $\mbox{U}(N)_k\times
\mbox{U}(N)_{-k}$ becomes $\mbox{U}(N+M)_k\times \mbox{U}(N)_{-k}$.
 To derive  the first part of  the
action (\ref{sc}),  we  take   just the  U(1) abelian   part of
$\mbox{U}(M)_{k}$ and assume that   the remaining  gauge  factors  are
spectators. Indeed,  on the
  D6-brane  one  can write  down  the following  $S_{WZ}$  action
\begin{equation}
S_{WZ}\sim T_{6}\int_{\mathbb{R}^{1,6}} F\wedge F\wedge A^{RR}_{3},
\end{equation}
 where $T_{6}$ is the D6-brane tension and where the gauge field  $A^{RR}_{3}$ is the R-R
3-form coupled to the D2-brane of type IIA superstring.  Integrating
by parts and  integrating  the result  over the  4-cycle $C_4$,  one
gets
\begin{equation}
-\frac{k}{4\pi} \int_{\mathbb{R}^{1,2}} A\wedge F
\end{equation}
where $k=\frac{1}{2\pi}\int_{C_4}(dA^{RR}_3)$  is  produced  now  by
$k $ D4-flux. To couple the system to an external gauge field, we
need to turn on the RR 5-form $A^{RR}_{5}$ which is coupled to the
D4-brane. Decomposing this gauge field  as follows
\begin{equation}
A^{RR}_{5}\to\tilde A \wedge\omega
\end{equation}
where $\omega$ is a harmonic 4-form  dual to  the four-cycle $C_4$,
the WZ  action $\int_{\mathbb{R}^{1,6}}
 A^{RR}_{5} \wedge F$ on
a D6-brane gives the second  term of the action (\ref{sc}), namely,
\begin{equation}
q \int_{\mathbb{R}^{1,2}} \tilde A \wedge F
\end{equation}  where    $q=\int_{C_4}w$ and  $\tilde A$  is an  U(1)
gauge field which   serves as the external gauge field that couples to the
gauge fields living on  the D6-brane world-volume. The   above
effective action on the D6-brane  world-volume  reproduces   the following filling factor
\begin{equation}
\label{nu}
\nu=\frac{q^2}{k}.
\end{equation}
From this equation, it follows  that the filling factor depends on
the  D4-branes and  the  harmonic 4-forms  defined on $\bf CP^3$. It
turns out that  known values could  be obtained  by taking
particular choices of such parameters. Moreover, these  FQHS in
AdS$_4$/CFT$_3$ have been extended to models  based on IIA dual
geometry realized as the  blown up of $\bf CP^3$ by four-cycles
which are isomorphic to $\bf CP^2$.  In particular, we  proposed a
stringy hierarchical description  of multi-layer systems in terms of
wrapped
D6-branes on the blown up four-cycles \cite{B}.\\
In what  follows,  we  investigate  FQHS system embedded  in CS
quiver gauge theories arising from M-theory  on  the cotangent
bundle on a two dimensional complex   toric variety $\bf V^2$. More
precisely,  we discuss the case of  two dimensional complex
projective space and the zeroth  Hirzebruch surface.

\section{FQHS and Chern-Simons quivers}
Many models  have been given to extend     ABJM theory.  They
are conjectured  to be    gauge field   duals    of  $AdS_4$
background  in type IIA   superstring and
   M-theory compactifications on eight dimensional manifolds.  In particular, these kinds of
      CS quiver
theories can  be obtained    in terms of M2-branes placed at  hyper
toric singularities of eight-dimensional manifolds. Recall that, the
simple model is described by  abelian gauge factors
$\mbox{U}(1)_{k_1}\times \mbox{U}(1)_{k_2}\times \ldots \times
\mbox{U}(1)_{k_n}$,  where $ k_i$ denote  the  Chern-Simons levels
for each abelian   factor $\mbox{U}(1)$. Geometrically, these models
can be encoded in  a quiver   formed  by $n$ nodes where each factor
$\mbox{U}(1)$ is associated  with a node   while the matter is
represented by the link  between nodes \cite{MS,BDGS}.   For these
models, it has been shown that Chern-Simons levels $ k_i$ are
subject to
\begin{equation}
\label{con} \sum_ik_i=0.
\end{equation}
 For the  $ \prod_i\mbox{U}(N_i)_{k_i}$  non abelian   gauge group, the Chern-Simons levels  $k_i$
  and the set of the ranks of the gauge groups $N_i$  are constrained by
\begin{equation}
\label{cong} \sum_ik_iN_i=0.
\end{equation}
The  CS quivers we consider here are obtained from M-theory on the
cotangent bundle over a two dimensional toric variety ${\cal \bf
V^2}$. This internal space is built in terms of a bi-dimensional
U(1)$^{r}$ sigma model with eight supercharges ($\mathcal{N}=4$) and
$r+2$ hypermultiplets. It has been considered to be  the solution of
the following D-flatness condition
\begin{equation}
\sum_{i=1}^{r+2}Q_{i}^{a}[\phi _{i}^{\alpha }{\bar{\phi}}_{i\beta
}+\phi _{i}^{\beta }{\bar{\phi}}_{i\alpha
}]=\vec{\xi}_{a}\vec{\sigma}_{\beta }^{\alpha },\;\; a=1,\ldots,r
\label{sigma4}
\end{equation}
where $Q_{i}^{a}$ is a matrix charge specified later on.  $\phi
_{i}^{\alpha }$'s ($ \alpha=1,2)$ denote the  component field
doublets of each hypermultiplets ($ i=1\ldots,r+2)$. $\vec{\xi}_{a}$
are  the   Fayet-Illiopoulos (FI)  3-vector couplings rotated by the
SU(2) symmetry, and $\vec{\sigma}_{\beta }^{\alpha }$ are the traceless $%
2\times 2$ Pauli matrices.   The
 explicit  solution   depends on the values of the FI
   couplings.  For  a particular region in the moduli space where   $\xi^1_a=\xi^2_a=0$  and
  $\xi^3_a >0$,  (\ref{sigma4}) describes the
  cotangent bundle over a  complex  two-dimensional toric variety  $\bf V^2$  defined by
\begin{equation}
\label{sigman2}
  \sum\limits_{i=1}^{2+r} Q_i^a|\phi^1_i|^2 = \xi^3 _a, \quad
a=1,\ldots,r.
\end{equation}
Using  M-theory/type IIA duality, it follows that M theory
compactified on such geometries  is dual to type IIA superstring on
3-dimensional complex manifolds $X_3$ fibered over a real line $R$
with D6-branes. In fact,  one   can  show   that  $X_3$ can be
described as a $\bf CP^1$ fibration over the  base $\bf V^2$.
Instead of being general, let us consider  first a toy model  where
$\bf V^2$ is
     $\bf CP^2$  associated with  $r=1$ and the vector charge $Q_i=(1,1,1)$.   The  corresponding
         quiver   has  3  nodes, where each
      node  is associated with an  U($N$) gauge  symmetry factor. For this CS quiver,
      we have a level  vector $(k_1,k_2,k_3)$ such that
 \begin{equation}
\label{concp2} k_1+k_2+k_3=0.
\end{equation}
It turns our that this model  could  be related to  ABJM theory   by
adding D6-branes wrapping 4-cycles of the  blown up  $\bf CP^3$  by
$\bf CP^2$   at singular toric points. In this way of  thinking and
solving the constraints  (\ref{cong}), the gauge symmetry
$\mbox{U}(N)_k\times \mbox{U}(N)_{-k}$ of ABJM theory
   can change into
$\mbox{U}(N)_k\times \mbox{U}(N)_{k} \times \mbox{U}(N)_{-2k}$. The
IIA/M-theory duality
 predicts that this  gauge  theory  should be dual  to type IIA superstring on  $AdS_4\times
 {\bf CP^1}\times {\bf CP^2}$. Roughly speaking,  the   CS gauge theory describing  FQHS   model
can be obtained from D4-branes   moving on  ${\bf CP^1}\times {\bf
CP^2}$. In type IIA geometry,  D4-branes can wrap $\bf CP^1$ and a
particular complex curve class $[C]$ in $H^2( {\bf CP^2},Z)$. On the
gauge theory side, the gauge symmetry becomes $\mbox{U}(N+M)_k\times
\mbox{U}(N)_{k}\times \mbox{U}(N+M')_{-2k}$.
 The constraint  (\ref{cong}) requires that
\begin{equation}
M=2N\qquad M'=N.
\end{equation}
As before, the CS quiver theory describing  our FQHS system   will
be in the   $\mbox{U}(2N)_k\times \mbox{U}(N)_{-2k}$ part of
$\mbox{U}(3N)_{k}\times \mbox{U}(N)_{k} \times \mbox{U}(2N)_{-2k}$.
Extracting an  $\mbox{U}(2)\times \mbox{U}(1)$,  we  can obtain  the
FQH   field theory  from   D4-branes wrapping individually $\bf
CP^1$ and   $C$. To see  that let us consider  first  the case of
the abelain
 part U(1) corresponding to $C$.
Indeed, on the 5-dimensional world-volume of each D4-brane we have U(1)
gauge symmetry. The corresponding effective  action  can take the following form
\begin{equation}
 S_{D4}\sim T_4 \int_{\mathbb{R}^{1,2}\times C} d^5\sigma e^{-\phi}
 \sqrt{ -det (G+2\pi F)}+T_4\int_{\mathbb{R}^{1,2}\times C} F\wedge F\wedge
A_1^{RR}
\end{equation}
where $T_4$ is the brane tension and where the gauge field
$A_1^{RR}$ is the R-R 1-form coupled to the D0-brane of type IIA
superstring. Ignoring the first term and integrating by part, the WZ
action on the D4-brane world-volume becomes
\begin{equation}
\int_{\mathbb{R}^{1,2}\times C} F\wedge F\wedge A_1^{RR}=-\int_{\mathbb{R}%
^{1,2 }\times C} A\wedge F\wedge (dA_1^{RR})
\end{equation}
Then, we get the  well known  Chern-Simons term
\begin{equation}
S_{WZ}\sim \int_{\mathbb{R}^{1,2}} A\wedge F.
\end{equation}
The presence of the R-R gauge field sourced by the anti-D6-flux
leads to $-2k=\frac{1}{2\pi}\int_{C}(dA_1^{RR})$. To couple the
system to an external gauge field, one needs an extra D4-brane
wrapping the cycle $C$ in the presence of    the  RR 3-form
$A^{RR}_3$ which is sourced by a D2-brane dual  to a D4-brane. After
the compactification, this gauge field decomposes into
\begin{equation}
A^{RR}_3\to \tilde A \wedge \omega
\end{equation}
where $\omega$ is a harmonic 2-form on $C$. In this way, the WZ
action  $\int A^{RR}_3 \wedge F$ on a D4-brane  wrapping $C$ gives
\begin{equation}
q \int_{\mathbb{R}^{1,2}} \tilde A \wedge F
\end{equation}
where $\tilde A$ is the U(1) gauge field which can be obtained from
the dimensional reduction of the RR 3-form $A^{RR}_3$. This U(1)
gauge field can be interpreted as a magnetic external gauge field
that couples to our QHS. We can follow the same steps to construct a
non abelian effective Chern-Simons gauge theory with  U(2) gauge
fields. Roughly speaking, thanks to $\nu=\nu_1(\mbox{U(2)})
+\nu_2(\mbox{U(1)})$,  and putting the charge    like $q_i=(1,1,1)$,
the corresponding  filling factor  reads as
\begin{equation}
\nu=\frac{2}{k}- \frac{1}{2k}=\frac{3}{2k}.\label{factorb}
\end{equation}
\section{ FQHS in CS quivers   on  Hirzebruch surfaces}
Let us extend the  result obtained in previous sections  to quivers
with more than three  nodes.  There are various   ways of doing
that. One possibility  can be realized by replacing two dimensional
projective spaces either by the  Hirzebruch surfaces or the del
Pezzo surfaces. The corresponding CS quivers involve more than three
$\mbox{U}(N_i)$ gauge symmetry factors. The general study is beyond
the scope of the present work, though we will consider an  explicit
example corresponding to the  zeroth  Hirzebruch surfaces $F_0$.
Other examples may be
 dealt with in a similar manner.  We will briefly comment on various simple extension in the conclusion.\\
Recall from literature that   $F_0$  is  a two-dimensional toric
surface defined by a  trivial fibration of  $\bf CP^1$ over  $\bf
CP^1$.  In $N = 2$ sigma model language, $F_0$ is  realized as the
target space $\mbox{U(1)}\times \mbox{U(1)}$ gauge theory with four
chiral fields with charges $(1,1, 0,0)$  and $(0,0, 1,1)$.  As
proposed in \cite{BDGS}, the corresponding   CS  quiver  is  given
by $\mbox{U}(N)\times \mbox{U}(N)\times \mbox{U}(N)\times
\mbox{U}(N)$ where the CS levels are $(k,-k, k,-k)$. This model
corresponds to a circular quiver with four nodes. The FQHS system we
are interested in
 appears  as  a CS quiver model  obtained from   gauge theories living on
the world-volume of the wrapped D4-branes on  2-cycles in type IIA
geometry which is given by the  trivial fibration of $\bf CP^1$ over
$F_0$. Indeed, there are three complex curves  on which the
D4-branes can be wrapped.  D4-branes can wrap $\bf CP^1$
and two complex curves class $[C_i]$ in $H^2(F_0,Z)$. In this
realization of FQHS, the fractional D4-branes will give three gauge factors.
Indeed, consider three stacks $( M_1,M_2,M_3)$ of D4-branes.  On the
IIA gauge theory side, the gauge symmetry becomes
$\mbox{U}(N+M_1)_k\times \mbox{U}(N)_{-k}\times \mbox{U}(N+M_2)_{-k}
\times \mbox{U}(N+M_3)_{k}$. A particular solution of  (\ref{cong})
requires that
\begin{equation}
M_1=N, \qquad M_1=2N, \qquad M_3=N.
\end{equation}
Our  CS quiver theory describing  FQHS  systems   will be in the
$\mbox{U}(N)_k\times \mbox{U}(2N)_{-k} \times \mbox{U}(N)_{k}$ part.
As before, thanks to  $\nu=\nu_1(\mbox{U(1)}\times
\mbox{U(1)})+\nu_2(\mbox{U(2)})$,   we can find the  the filling
factor of the model. Indeed,  consider first   the  bi-layer system
corresponding to $\mbox{U}(1)\times \mbox{U}(1)$ quiver gauge
theory. The corresponding   matrix $K_{ij}$  takes the form
\begin{equation*}
K_{ij}(\mbox{U}(1)\times \mbox{U}(1))=\begin{pmatrix}
k & 0 \\
0& k%
\end{pmatrix}
\end{equation*}
 where the integer $k$  can   be interpreted  geometrically in  terms of the
   self intersection  of complex curves  with genus $
g>0$. From  type IIA string point of view, this model can be
obtained from two D4-branes wrapping  two  identical   complex
curves. Evaluating (\ref{factor}) for the charges $ q_{i}=(q_1,q_2)$
yields
\begin{equation}
\nu_1 =\frac{q_1^2+q_2^2}{k}.
\end{equation}
This is a general expression  depending on  the vector charge.
Particular choice of the D-brane flux and  the  charge vector may
recover some observed  values of the filling factor. Adding now the
non abelian contributions, the total filling factor can be written
as
\begin{equation}
\nu =\frac{q_1^2+q_2^2-2q_3^2}{k}.
\end{equation}
It  is worth noting that  the vanishing filling factor behavior has
to do with the  condition  $q_1^2+q_2^2=2q_3^2$.  This can  be fixed
by the flux on  the world-volume D6-branes.  The Hall conductivity
is quantized in terms of the CS level being identified with the
D6-flux and the charges given in terms of the integral of harmonic
2-forms over dual 2-cycles  embedded in type IIA geometry.
\section{ Conclusion}
In this letter, we have  studied  a class of  FQHS  in  toric CS
quivers from M-theory on the contangent bundle over a  two dimensional toric
 variety  ${\bf  V^2}$. The dual type IIA geometries are   realized as $AdS_4\times
 {\bf CP^1}\times \bf V^2$.
Using M-theory/Type IIA duality, we  have presented hierarchical
stringy descriptions using quiver gauge models living on  stacks of
D4-branes wrapping  2-cycles in  ${\bf CP^1}\times \bf V^2$  and
 interacting with  R-R gauge fields.
In particular,  we have given   two  simple examples of FQHS systems. The first model is  based
 on  the CS quiver  relying on  ${\bf CP^1}\times \bf CP^2$, while the second  one is associated with
quiver gauge theory describing
  system  based on  D4-branes  which wrap   2-cycles inside ${\bf CP^1}\times F_0$.
 This analysis can be  adapted  to  other multilayered  systems
   by considering   more general  two dimensional toric manifolds.
   The manifolds are given by  $ {\bf C}^{r+2}/ {{\bf C}^*}^r$,
   where the   ${{\bf C}^*}^r$ actions are  given by $ z_i \to \lambda^{q_i^a}
   z_i,\;\;i=1,\ldots, r+2; a=1,\ldots, r$.  Clearly, type IIA geometry
will have $h^{1,1}=r+1$. The right
interpretation of this is that we  have $r+1$  2-cycles associated  with  wrapped D4-branes.
 The corresponding  FQHS model involves  $r+1$  gauge factors. In connection with that, it would
therefore be of interest to consider such systems using  tools
developed  in toric geometry \cite{Fu,LV}. It is natural to consider
del Pezzo surfaces. Other possibilities  for $\bf V^2$ are local ALE
spaces with ADE singularities.

On the other hand, motivated by  supersymmetric part of ABJM theory and supersymmetric QHE, it should be interesting to study such  realizations
in terms of M-theory on  eight dimensional spaces. We hope, all these open questions will be addressed in
future works.

\hspace*{-15pt}

{\bf \emph{Acknowledgments}}:    The author would like to thank his
mother    for  patience.  He thanks also his collaborators for
discussions on related topics.


\begin{thebibliography}{99}
\bibitem {BBST}
    B. A. Bernevig, J. Brodie, L. Susskind and N. Toumbas,  {\em How Bob Laughlin Tamed the Giant
Graviton from Taub-NUT space}, JHEP{\bf 0102}(2001)003, {\tt  hep-th/0010105}.

\bibitem {FLRT}M. Fujita, W. Li, S. Ryu, T. Takayanagi, {\em Fractional
Quantum Hall Effect via Holography: Chern-Simons, Edge States, and
Hierarchy}, {\tt arXiv:0901.0924[hep-th]}.

\bibitem {BJLL}
O. Bergman, N. Jokela, G. Lifschytz, M. Lippert, {\em Quantum Hall
Effect in a Holographic Model}, {\tt arXiv:1003.4965[hep-th]}.


\bibitem {BS}A. Belhaj, A. Segui, \emph{Engineering of Quantum Hall Effect
from Type IIA String Theory on The K3 Surface}, Phys. Lett. \textbf{B691}%
(2010)261-267, {\tt arXiv:1002.2067[hep-th]}.

\bibitem {BFSS}
 A. Belhaj, N-E. Fahssi, E.H. Saidi, A. Segui, {\em  Embedding Fractional
 Quantum Hall Solitons in M-theory Compactifications}, {\tt arXiv:1007.4485}.
\bibitem {BEFKSS}
A. Belhaj, A. ElRhalami, N-E. Fahssi, M. J. I. Khan, E. H. Saidi, A.
Segui, {\em Brane Realizations of Quantum Hall Solitons and
Kac-Moody Lie Algebras}, {\tt arXiv:1008.0351 [hep-th]}.


\bibitem{ABJM} O. Aharony, O. Bergman, D.L. Jafferis, J. Maldacena, {\em
$N=6$ superconformal Chern-Simons-matter theories, M2-branes and
their gravity duals},
   JHEP {\bf 0810}(2008)091, {\tt  arXiv:0806.1218 [hep-th]}.



\bibitem {F}  M.  Fujita, {\em  M5-brane Defect and QHE in AdS4 $\times$ N(1,1)/N=3
SCFT}, Phys. Rev. {\bf D83}(2011)105016, {\tt
arXiv:1011.0154[hep-th]}.

\bibitem {HLT} Y. Hikida, W. Li, T. Takayanagi,  {\em ABJM with
Flavors and FQHE}, JHEP {\bf 0907}(2009)065, {\tt  arXiv:0903.2194}.



\bibitem {B}A. Belhaj, \emph{ On Fractional Quantum Hall Solitons in ABJM-like
Theory}, Phys. Lett. \textbf{B705}(2011)539-542 , {\tt
arXiv:1107.2295[hep-th]}.
\bibitem {GW}
D. Gaiotto, E. Witten, Janus Configurations, {\em  Chern-Simons Couplings, And The
theta-Angle in N=4 Super Yang-Mills Theory}, JHEP 1006(2010)097,{\tt arXiv:0804.2907
[hep-th]}.


\bibitem {L}R. B. Laughlin, \emph{Anomalous Quantum Hall Effect: An
Incompressible Quantum Fluid with Fractionally Charged Excitations},
Phys. Rev. Lett. \textbf{50}(1983)1395.

\bibitem {W}X-G. Wen, \emph{Quantum Field Theory of Many-body Systems},
Oxford University Press, 2004,

\bibitem {WZ}
 X.G. Wen, A. Zee, {\em Classification of
Abelian quantum Hall states and matrix formulation of topological
fluids}, Phys. Rev. {\bf B 46}(1992)2290-2301.




\bibitem{MS}
 D. Martelli, J. Sparks, {\em Moduli spaces of Chern-Simons quiver gauge theories and $AdS_4/CFT_3$},
 Phys.Rev.{\bf D78}(2008)126005, {\tt  arXiv:0808.0912 [hep-th]}.


\bibitem{BDGS} A.
Belhaj, P. Diaz, M. P.  Garcia del Moral, A. Segui,  {\em  On
Chern-Simons Quivers and Toric Geometry}, {\tt
arXiv:1106.5335[hep-th]}.

\bibitem {Fu}W. Fulton, {\em Introduction to Toric Varieties}, Annals of Math.
Studies, No. \textbf{131}, Princeton University Press, 1993.

\bibitem{LV}
N.C. Leung and C. Vafa, {\em Branes and Toric Geometry},   Adv.
Theo. Math. Phys {\bf 2}(1998) 91, {\tt hep-th/9711013}.


\end{thebibliography}
\end{document}